%% file: main.tex
\documentclass[sigconf,nonacm]{acmart}
\usepackage{caption}
\usepackage{subcaption}
\usepackage{svg}
\usepackage{hyperref}
\usepackage{enumitem}
\usepackage{multirow}

\newcommand{\revise}[1]{\textcolor{black}{#1}}

\AtBeginDocument{%
  }

\setcopyright{none}                     
\acmDOI{} \acmISBN{} \acmPrice{} \acmYear{} \copyrightyear{}
\acmConference[]{}{}{} \acmBooktitle{}

\usepackage{multicol}
\usepackage{graphicx}
\usepackage{multirow}
\usepackage{algorithm}
\usepackage{cuted}
\usepackage{algcompatible}
\usepackage{enumitem}
\usepackage{xspace}

\newcommand\dbname{\ensuremath{\textsf{Aixel}}\xspace}

\begin{document}

\title{\dbname: A Unified, Adaptive and Extensible System for AI-powered Data Analysis}

\author{Meihui Zhang}
\email{meihui_zhang@bit.edu.cn}
\orcid{0000-0002-0752-9877}
\affiliation{%
  \institution{Beijing Institute of Technology}
  \city{Beijing}
  \country{China}
}

\author{Liming Wang}
\email{liming_wang@bit.edu.cn}
\orcid{0000-0002-0109-5668}
\affiliation{%
  \institution{Beijing Institute of Technology}
  \city{Beijing}
  \country{China}
}

\author{Chi Zhang}
\email{chi_zhang@bit.edu.cn}
\orcid{0009-0000-0323-0715}
\affiliation{%
  \institution{Beijing Institute of Technology}
  \city{Beijing}
  \country{China}
}

\author{Zhaojing Luo}
\email{zjluo@bit.edu.cn}
\orcid{0000-0001-7271-3999}
\affiliation{%
  \institution{Beijing Institute of Technology}
  \city{Beijing}
  \country{China}
}

\renewcommand{\shortauthors}{Meihui Zhang et al.}

\begin{abstract}
\revise{A growing trend in modern data analysis is the integration of data management with learning, guided by accuracy, latency, and cost requirements.}
In practice, applications draw data of different formats from many sources.
In the meanwhile, the objectives and budgets change over time.
Existing systems handle these applications 
across databases, analysis libraries, and tuning services.
\revise{Such fragmentation leads to complex user interaction, limited adaptability, suboptimal performance, and poor extensibility across components.}
To address these challenges, we present \dbname{}, a unified, adaptive, and extensible system for AI-powered data analysis.

\revise{The system organizes work across four layers: application, task, model, and data.
The task layer provides a declarative interface to capture user intent, which is parsed into an executable operator plan.
An optimizer compiles and schedules this plan to meet specified goals in accuracy, latency, and cost.
The task layer coordinates the execution of data and model operators, with built-in support for reuse and caching to improve efficiency.
The model layer offers versioned storage for index, metadata, tensors, and model artifacts. It supports adaptive construction, task-aligned drift detection, and safe updates that reuse shared components.
The data layer provides unified data management capabilities, including indexing, constraint-aware discovery, task-aligned selection, and comprehensive feature management.}
With the above designed layers,
\dbname{} delivers a user friendly, adaptive, efficient, and extensible system.
\end{abstract}






\maketitle

\input{sections/1-intro}
\input{sections/2-background}

\input{sections/3-overview}

\input{sections/4-design-data}

\input{sections/5-design-model}
\input{sections/6-design-task}
\input{sections/7-conclusion}

\bibliographystyle{ACM-Reference-Format}
\bibliography{system.bib}
\end{document}

%% file: sections/1-intro.tex
\section{Introduction}
\label{sec:intro}

\revise{
Modern data analysis increasingly couples data management with learning under accuracy, latency, and cost constraints.
Although the community has introduced many advances in database systems, analytical tools, and model optimization methods, existing solutions still face several challenges. 
}

Traditional database management systems (DBMSs) focus primarily on data storage, retrieval, and transactional guarantees~\cite{astrahan1976systemR,postgresql1996postgresql}, but they lack native support for managing or executing AI models. 
These systems treat models as external artifacts.
As a result, model training and inference remain decoupled from data execution paths, leading to redundant data movement, fragmented execution, and increased interaction overhead. 
For users, this fragmentation often translates into complex, multi-step orchestration across disconnected tools, making end-to-end data analysis
cumbersome and error-prone. 
These limitations motivate the need for a unified, user-friendly analysis system that reduces interaction overhead by tightly coupling data and model execution, while supporting adaptive, cost-aware processing within a shared framework.

\revise{Data-centric analysis tools (e.g., Pandas for data selection and transformation; TensorFlow and PyTorch for model training)} 
provide powerful primitives for data selection, transformation, and model training, enabling streamlined development across diverse analysis workflows
~\cite{mckinney2011pandas,developers2022tensorflow,imambi2021pytorch}. 
\revise{However, these tools lack unified support for indexing across heterogeneous data sources and offer no integrated scheduling across data and model operations~\cite{bradshaw2019mongodb,lakshman2010cassandra,vora2011hbase}.}
In practice, data access depends on external stores, while model tuning and deployment are handled separately~\cite{GolovinSMKKS17}, resulting in fragmented execution and increased coordination overhead. 
This disconnect limits the system’s ability to adapt to dynamic data scenarios and prevents holistic optimization under accuracy and cost constraints. 
These limitations motivate the design of a unified data analysis system that reduces manual orchestration, supports adaptive execution across varying data conditions, and enables extensibility through shared infrastructure.

Beyond general-purpose libraries, automated model optimization methods provide extended support for hyperparameter tuning, pipeline search, and deployment~\cite{FeurerEFLH22,HeffetzVKR20,LiJDRT17,FalknerKH18,GolovinSMKKS17}. 
However, these methods typically operate downstream from data processing and rely on curated inputs, assuming that data preparation has been handled externally. 
This separation introduces versioning inconsistencies, and complicates performance monitoring across data and model stages.  
In dynamic production environments, such decoupling delays response to data distribution shifts and makes it difficult to detect leakage or trace provenance, ultimately degrading accuracy and increasing operational cost~\cite{LiRBZCZ21,PhaniR021}. 
These challenges underscore the need for a unified system that integrates automated modeling with data management and jointly optimizes both under shared accuracy, latency, and cost constraints, while supporting adaptive responses to evolving data conditions.

Overall, existing methods do not provide a unified end-to-end system that integrates data, models, and tasks within a shared execution framework. 
Current approaches either separate these components or handle them independently, creating fragmented workflows that lack coherence and efficiency. 
\revise{
As a result, these trends motivate a data analysis system that reduces interaction overhead through a user-friendly design, supports adaptation to evolving data scenarios, maintains performance under accuracy and cost constraints, and enables strong extensibility.
}

Therefore, we propose \dbname{}, a unified platform that integrates data management, model management, and task orchestration into a single cohesive system.
\dbname{} treats data, models, and tasks as first-class entities, enabling seamless transitions across all stages.
The system design follows four core principles: user friendliness, adaptivity, efficiency, and extensibility.
It is designed to seamlessly support machine learning workflows, integrating AI models directly within the system's core.
\dbname{} enables the advancement of data, model and task, optimizing end-to-end workflows while maintaining flexibility, scalability, and efficient performance.
\revise{According to this , the core of \dbname{} comprises three integrated layers: the task layer, the model layer, and the data layer.
The task layer provides a user-friendly declarative interface that captures analytical intent and compiles it into executable plans. 
It orchestrates cross-layer execution and applies adaptive strategies to satisfy effectiveness and efficiency.
The model layer offers extensible support for model storage, versioning and management. 
It enables efficient training and maintenance workflows while reuse across tasks through shared components.
The data layer manages various data with unified indexing, constraint-aware discovery, and task-aligned data access. 
It supports intelligent data selection and rich feature processing, enabling scalable and efficient downstream learning.
Through this design, \dbname{} delivers an end-to-end system from user intent to optimized execution—aligning data and models under a unified, adaptive, and extensible framework.}
We also outline future directions for the improvement of \dbname{}, highlighting the research opportunities that can enhance its efficiency, adaptability, and integration with emerging AI techniques.

%% file: sections/2-background.tex
\section{Background}
\label{sec:background}

\subsection{Development of Database Management Systems}
The development of database management systems (DBMS) proceeds through several distinct phases, each driven by evolving data types, scalability requirements, and application demands. 
Relational databases such as System R~\cite{astrahan1976systemR} and PostgreSQL~\cite{postgresql1996postgresql} establish a foundation for structured data with SQL, schema enforcement, and ACID (atomicity, consistency, isolation, durability) guarantees, and remain dominant in enterprise and OLTP workloads.

To support large-scale web data, NoSQL systems~\cite{bradshaw2019mongodb, lakshman2010cassandra, vora2011hbase} (e.g., MongoDB, Cassandra, HBase) offer greater scalability and schema flexibility.
These systems relax consistency in favor of availability and partition tolerance, support horizontal scaling, and adopt diverse data models including document, key-value, and wide-column stores.
However, the lack of declarative querying and strong guarantees limits their adoption in mission-critical analysis. 

To address this gap, NewSQL systems~\cite{corbett2013spanner, taft2020cockroachdb, huang2020tidb} (e.g., Google Spanner, CockroachDB, TiDB) combine the scalability of NoSQL with the transactional semantics and SQL support of traditional DBMS.
These systems rely on distributed consensus protocols and global timestamps to maintain consistency at scale, and see widespread adoption in globally distributed transactional applications.

Most recently, AI × DB systems aim to natively support machine learning workflows within databases. For example, NeurDB~\cite{ooi2024neurdb} integrates deep neural models into the query execution layer, embeds inference into physical plans, and enables hybrid reasoning over structured data and neural representations.

Different from DBMS, \dbname{} is not limited to data management.
It natively manages models, including storage, versioning, construction, and maintenance, and exposes a user-friendly interface for invoking diverse tasks through a declarative workflow.
Built as an AI-powered engine, \dbname{} couples a unified data index with model-aware operators and an execution optimizer that reasons over both data and models.
\dbname{} enables end-to-end planning, unified governance, and efficient execution, distinguishing \dbname{} from systems that attach learning components as external services.

\subsection{Data-Centric Analysis Tools}
Data-centric analysis tools focus on in-memory computation over tables and tensors, and provide the core building blocks for data preparation, feature engineering, and model training.
Pandas~\cite{mckinney2011pandas} offers labeled tabular structures and a rich toolkit for selection, joins, grouping, reshaping, and time-series handling, making it the de facto environment for data cleaning and feature crafting on a single machine.
TensorFlow~\cite{developers2022tensorflow} provides a graph-based computation framework with automatic differentiation and optimized runtime for large-scale model training and deployment.
PyTorch~\cite{imambi2021pytorch} emphasizes intuitive, dynamic computation and supports both experimentation and production through tracing, compilation, and distributed training.
Together, these systems enable efficient workflows from data transformation to model training. 
However, they focus on isolated computation and rely on external systems for data management, indexing, and execution optimization. 

In contrast, \dbname{} offers an integrated environment that unifies data management and model execution within a single system.
Its execution engine enables seamless coordination between data, models, and tasks under a unified layer, allowing smooth transitions from data processing to model inference.
By integrating learned representations directly into the pipeline, \dbname{}  supports coherent, end-to-end workflows within a consistent execution framework.
This unified design eliminates fragmentation across stages, streamlines the development of data-driven applications, and positions \dbname{}  as a cohesive foundation for building intelligent, analysis-integrated systems.

\subsection{Automated Model Optimization}
Work on automated model optimization has moved beyond simple model selection and hyperparameter tuning to richer automation over the construction and search loop. 
Meta-learning portfolios and bandit scheduling reduce manual effort and warm-start search in systems such as~\cite{FeurerEFLH22}, while reinforcement learning generates end-to-end pipelines with operator choices as in~\cite{HeffetzVKR20}. 
Cost control is addressed by multi-fidelity and hybrid optimization methods like~\cite{LiJDRT17,FalknerKH18}, and production services such as~\cite{GolovinSMKKS17}. 
Learning systems also improve throughput and deployment efficiency via data- and compute-aware selection~\cite{NakandalaZK21}, approximate training with guarantees~\cite{ParkQSM19}, and hardware-aware specialization~\cite{CaiGWZH20}.

Despite these advances, most approaches assume that inputs are already curated and that downstream deployment can be handled separately. 
Empirical studies show that data quality shifts outcomes and should be considered with model choice ~\cite{LiRBZCZ21}. 
Systems for repair and detection provide useful mechanisms, for example probabilistic cleaning~\cite{RekatsinasCIR17} and configuration-free error detection~\cite{MahdaviAFMOS019}, and recent work optimizes cleaning pipelines for ML utility~\cite{SiddiqiKB23} and captures fine-grained lineage for reuse and leakage control~\cite{PhaniR021}. 
However, these pieces are usually external to automated analysis planners, and they rarely co-optimize data discovery, task-aligned selection, provenance, and deployment constraints together with model construction and maintenance.

\dbname{} addresses this gap by treating automated analysis as an operator-based workflow that unifies data management.
Within a single plan, data indexing, discovery, and task-aligned selection are coordinated with model construction, drift detection, and update, so that data quality, leakage control, and provenance directly influence which predictors are used and how they evolve.
A cost-aware optimizer balances accuracy, latency, and resource use end to end, and a versioned model store preserves the context needed for audit and refresh.
Unlike prior work that treats data preparation and model search as separate concerns, \dbname{} provides a single stack that is easy to use, adaptive to change, and extensible.

%% file: sections/3-overview.tex
\section{Overview}
\label{sec:overview}
In this section, we present the key principles that guide the design of \dbname and set the stage for its conceptual and architectural overview.  

\subsection{Design principles}
\begin{figure}
\centering \includegraphics[width=\linewidth]{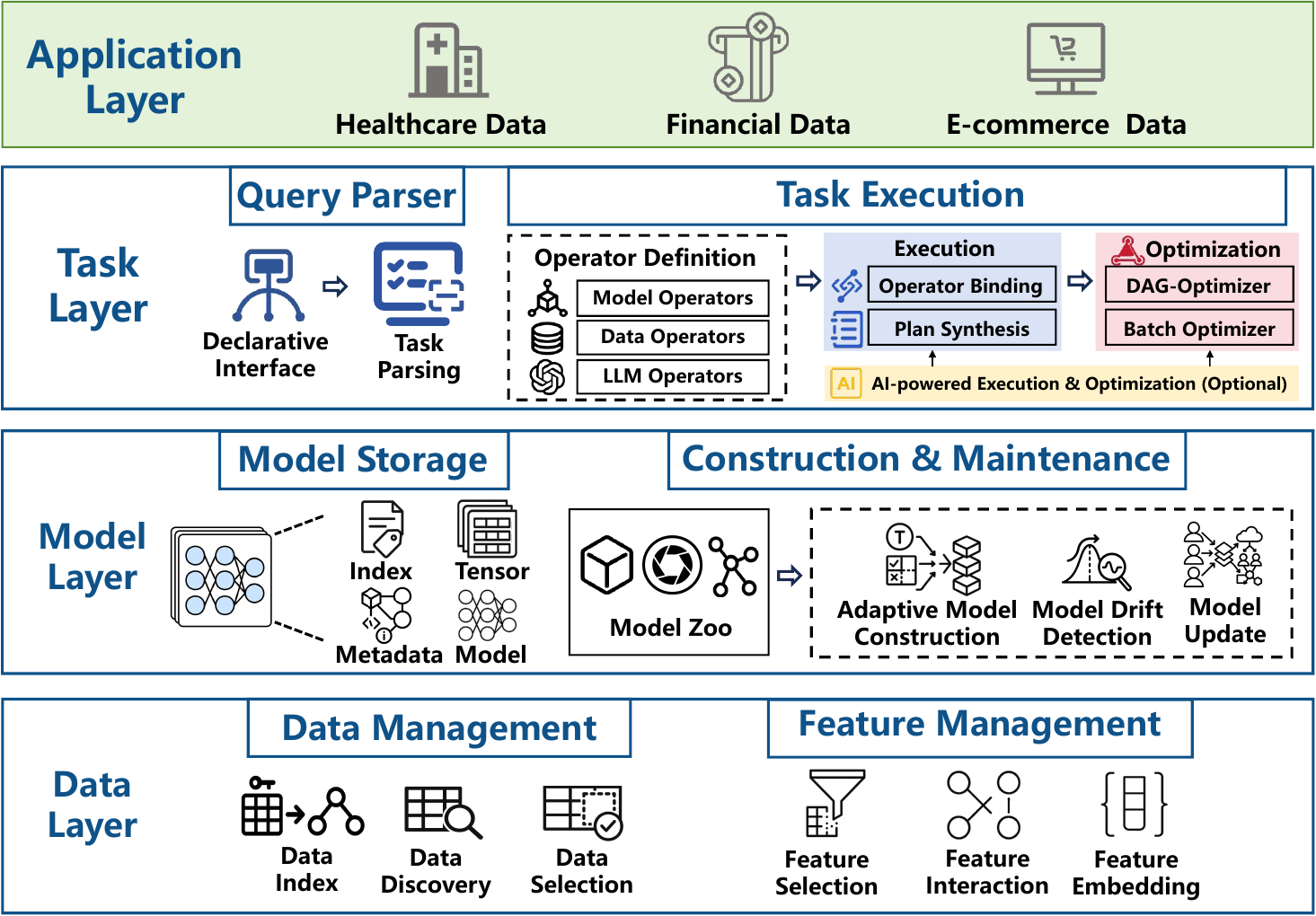} 
\caption{The System Architecture of {\dbname{}}} 
\label{fig:architecture} 
\vspace{-3mm}
\end{figure}

Modern analysis spans heterogeneous applications and evolving data, so a practical system must adapt during planning, learning, and execution.   
Traditional systems depend on manual reconfiguration and offline model updates, which delay response and weaken performance. 
In contrast, \dbname{} organizes analysis in a four-layer architecture (Figure~\ref{fig:architecture}) comprising the application, task, model, and data layers. 
From this vantage, we distill four design properties that guide the system: user friendliness, adaptivity, performance optimization, and extensibility.

\textbf{User friendliness.}
\dbname{} is designed with user friendliness as a core principle, ensuring that users can express analytical intent clearly without requiring deep system knowledge.
It provides intuitive interaction through both natural language and declarative interfaces, supporting flexible and efficient task formulation.
At the task layer, the AxielParser interprets user intent from these inputs and transforms it into an executable operator graph that connects data, model, and LLM components.
By abstracting away complex logical mappings, AxielParser significantly lowers the barrier for users to engage with AI-powered data analysis.
\dbname{} further improves usability by providing readable plan explanations and interactive previews at key stages within both the data and model layers, while also offering actionable feedback when inputs are incomplete or constraints conflict.
Together, these capabilities simplify complex workflows and make \dbname{} equally suitable for everyday analysts and experts in domains such as healthcare, finance, and e-commerce.

\textbf{Adaptivity.}
\dbname{} adapts to changing objectives and evolving data through coordination between the data layer, the model layer, and the task layer. 
At the data layer, adaptation addresses the dynamic nature of data. 
The system performs constraint-informed discovery and refreshes task-aligned selections as sources evolve. 
As schemas and distributions change, it maintains corresponding feature views. 
At the model layer, adaptation centers on the model lifecycle. 
\dbname{} constructs models tailored to the current data context and objective. 
It also performs task-aligned monitoring to detect drift. 
Furthermore, it manages versioned updates to permit comparison and rollback, all within bounded compute and memory resources.
At the task layer, the system focuses on adaptive execution.  
It compiles intent into plans that adhere to stated budgets.  
In response to observed statistics and changing constraints, it also adjusts operator bindings, batch sizes, and degrees of parallelism.  
These adaptations are performed without altering the application interface.

\textbf{Efficiency.}
To ensure effectiveness and efficiency across diverse workloads, \dbname{} incorporates performance optimization techniques at every stage of execution.
At the task layer, a task optimizer explores multiple plan candidates using a DAG-based planner and batch-level optimization, optionally guided by learned cost models.
Data operators and feature pipelines are fused to minimize redundant scans and intermediate materialization.
Index structures and cached tensors are aligned with the selected execution plan to accelerate repeated access.
Model operators leverage batching and lightweight inference paths, while admission control and micro-batch shaping mechanisms balance throughput and latency.
These techniques work together to deliver predictable speedups without compromising on accuracy or reproducibility.

\textbf{Extensibility.}
\dbname{} is built from modular and versioned components so that capabilities can evolve safely. 
Operator Definition exposes stable interfaces for data, model, and LLM operators, and each operator declares capabilities and constraints for correct composition. 
Model storage separates index, metadata, tensor, and model artifacts, which simplifies upgrades and side-by-side evaluation. 
Policies in construction and maintenance are configurable, allowing domains to register new drift signals, update rules, and specialized embeddings without redesign. 
Clear contracts across layers enable incremental growth of the system.

\subsection{System Architecture Overview}
This subsection summarizes the main components of \dbname{} and how they cooperate across layers.
The design follows the framework in Figure~\ref{fig:architecture}.
Each layer defines stable interfaces that link query parsing, operator definition, operator binding and plan synthesis, optimization, model storage and maintenance, and data and feature management into a coherent execution workflow.

\textbf{Application layer.}
\dbname{} targets diversified AI data analysis tasks.
Applications express analysis intent through a declarative interface, while the system provides consistent execution, auditable decisions, and reusable artifacts without exposing operator or model management to end users.

\textbf{Task layer.}
The task layer turns user intent into an executable plan.
The query parser accepts a declarative request and produces an operator graph with explicit inputs, dependencies, and expected outputs.
Task execution grounds this graph by binding operators to concrete implementations and synthesizing a runnable DAG.
Optimization explores plan alternatives with a graph optimizer and a batch optimizer.
When beneficial, learned cost models guide selection among competing choices.
The layer emits the final plan together with metadata for monitoring and explanation.

\textbf{Model layer.}
The model layer manages models and their artifacts.
Model Storage separates four artifact types: index, metadata, tensor, and model binaries.
This separation supports efficient lookup, lineage, and reuse.
Construction and Maintenance governs the evolution of predictors through a model zoo, adaptive model construction, drift detection, and versioned updates.
These mechanisms allow the system to react to new objectives and data while keeping changes safe.

\textbf{Data layer.}
The data layer prepares inputs for execution and produces meta-features that inform routing and plan choices.
Data Management maintains data indexes, supports dataset discovery, and selects task-specific subsets when needed.
Feature Management carries out feature selection, feature interaction, and feature embedding.
Its outputs feed both the task plan and the model layer, and cached results are reused across requests to reduce cost.

%% file: sections/4-design-data.tex
\section{Data Layer}
\label{sec:data}

\subsection{Data Management}
The Data Layer in \dbname{} focuses on data Index and data Discovery. 
Its goal is to make heterogeneous data easy to find, fast to retrieve, and safe to use for multiple tasks.
It provides consistent metadata, efficient indexing, and task aware retrieval while honoring various constraints.

\subsubsection{Data Index}
Data indexing is fundamental to efficient discovery in \dbname{}. 
Unlike traditional indexes\cite{malkov2018HNSW, yang2024vdtuner} for vectors that optimize a single similarity objective and often ignore structured constraints, \dbname{} focuses on a fusion graph index that unifies vector similarity with interval and label constraints and remains robust across heterogeneous data types.

In \dbname{}, we introduce AixelIndex, a fusion graph index that unifies vector similarity with interval and label constraints for heterogeneous data types.
Technically, AixelIndex models each object as a node with an embedding, a sortable numeric attribute, and tags. 
Edges encode interval and label metadata, and the graph is built by incremental insertion with local neighbor search and dominance pruning, partitioned by time and tenant, and supported by online statistics for selectivity. 
The index stores compressed adjacency with cache friendly ordering to speed traversal and reduce memory.
AixelSearch cuts redundant traversal and accelerates query time while raising precision, especially under strong constraints. In deployment, it produces smaller, higher-quality candidate sets for the execution optimizer, improving efficiency.

\subsubsection{Data Discovery}
Data Discovery is essential to deliver low latency and accurate results in \dbname{}. 
Traditional discovery pipelines first run a broad candidate search with text or vector indexes and then apply filters, which wastes computation on irrelevant candidates and degrades precision when range and label constraints are strong. 

Our key insight is to embed constraints into traversal, limiting exploration to a query specific subgraph. 
Building on AixelIndex, we propose the AixelSearch algorithm.
Concretely, AixelSearch first derives a constraint profile from the decomposed intent, including the numeric range and the label set. 
It then prunes edges that violate the range or have no label overlap, yielding a compact working subgraph before any heavy scoring is performed. 
Within this subgraph, AixelSearch performs guided neighbor expansion that interleaves similarity checks with explicit constraint filters for ranges and labels, and it prioritizes nodes whose metadata indicate high evidence density. 
A lightweight ranker aggregates signals from similarity, range fit, and label coverage to produce a final candidate set and optional evidence slices for downstream operators.

AixelSearch reduces unnecessary traversal, improves precision under strong constraints, and shortens end to end latency. 
In practice, AixelSearch provides the execution optimizer with smaller and higher quality candidate sets, which translates into better answer quality and lower token and compute budgets.

\subsubsection{Data Selection}
In practice, available data are large, mixed in quality, and change over time.
A single task may see many datasets or partitions, with duplicated rows, missing fields, and stale entries.
Data Selection identifies a task-relevant, representative working set from the discovered sources.
The objective is to align the data with the analysis goal (e.g., target variable, evaluation metric, latency budget) so that models train and infer on examples that matter for the task.

Our key insight is to select data by task utility.
In \dbname{}, we keep the process simple to read and apply, adaptive to data shifts, efficient before heavy processing.
Concretely, \dbname{} uses lightweight statistics and a decomposed task intent to prefer diverse, well-formed samples that cover key slices of the task, while down-weighting or excluding redundant and low-signal records.  
By focusing computation on a compact and representative working set, downstream feature pipelines and models perform fewer scans and less materialization, which shortens runtime and lowers resource usage.
When sources evolve, the same task-aligned selection criteria can be refreshed to maintain comparability across runs, without changing the analysis interface. 
Finally, selection separates training, validation, and serving segments to reduce leakage and keep evaluation reliable.

\subsection{Feature Management}
Feature Management converts the task aligned working set into features that models can use. 
It organizes this process into three parts: Feature Selection, Feature Interaction, and Feature Embedding. 
The module connects inputs from Data Selection with the needs of the Task Layer and the Model Layer.
The goal is to produce features that match the stated target, evaluation metric, and latency goal.

\subsubsection{Feature Selection}
In practice, a single analysis task often draws on large, evolving data sources with uneven quality, correlated signals, and missing values. 
Passing all attributes to the model raises cost, slows planning, and can harm generalization. 
Feature selection determines which attributes from the task aligned working set are passed to the model.
This selection is challenging for several reasons.  
High-dimensional inputs make exhaustive search infeasible under tight budgets.  
Strong correlations create redundant signal and lead to unstable importance estimates.  
Distributions shift over time and across organizations or accounts, so a choice that works on one dataset may degrade on another.  
In collaborative settings across organizations, attributes are split across parties, which introduces privacy and communication constraints.

Our central idea is to align selection with task utility rather than broad inclusion, and to refresh choices when data shift without rewriting pipelines.
To realize this plan, \dbname{} proposes methods for common settings. 
For rapid decisions in single-organization settings, we propose PA-FEAT~\cite{PAFEAT}, which transfers selection knowledge from prior tasks and applies progress-aware multi-task reinforcement learning to produce high-quality subsets under tight time budgets. 
For cross-organization settings with vertically partitioned attributes, we propose FEAST~\cite{FEAST}, which scores candidates with conditional mutual information to retain informative yet nonredundant attributes and uses communication-efficient coordination that preserves data boundaries.
These methods provide a unified, task-aligned selection capability that is easy to apply.

\subsubsection{Feature Interaction}
Many analysis tasks require features to be considered together because useful signals emerge only when they interact. 
Ignoring such interactions can hide patterns that matter for the target and metric, while constructing all possible combinations creates excessive candidates, raises cost, and increases the risk of overfitting. 
This is challenging because the number of potential combinations grows quickly, noisy or imbalanced data can create spurious relations, and schemas and source distributions change over time. 

We build interaction features in a task aligned and evidence driven manner. 
Candidates are suggested from simple association signals, recent errors of a lightweight baseline, and available knowledge, and only those that show consistent utility under the current plan are retained.
In \dbname{}, we propose methods that focus on different aspects of data feature interaction modeling.
Att-Reg~\cite{AttReg} models pairwise relations between features and adjusts their strength according to usefulness, which reduces overfitting and improves stability. 
PFCA~\cite{PFCA} filters paths from external knowledge and emphasizes joint effects that are tied to the outcome, which is helpful when auxiliary links are available. 
DMRNet~\cite{DMRNet} combines current information with important links from history and keeps those that remain predictive. 
ARM-Net~\cite{ARMNet} selects and weighs cross features adaptively so that helpful combinations are kept and noisy ones are suppressed. 
LDA-Reg~\cite{LDAReg} uses summaries from text to guide interaction strength when textual descriptions are present. 
An adaptive regularization method~\cite{GMReg} further adjusts penalties based on observed parameters, which stabilizes selected interactions under heterogeneous inputs.
These methods are provided through a unified interface that selects a small set of useful feature combinations for the current task.  
The outcome is a compact interaction view that improves analysis quality and controls cost.

\subsubsection{Feature Embedding}
Many analysis tasks combine signals from numeric fields, categorical codes, free text, and relations across sources.  
Naive encodings lose meaning, while heavy encoders can exceed latency and memory limits.  
As a result, models may see weak or incomplete signals and plans may slow down.  
The immediate need is to form compact representations that preserve useful semantics and relevant context without breaking performance budgets.
This goal is difficult because inputs are always heterogeneous, contextual cues vary in importance, and embeddings must be refreshed as data evolve.  
A single embedding method rarely suits all attribute types, combining neighbor signals without selection can introduce noise, and recomputing all embeddings from scratch is costly.

We make embedding semantics driven and aligned with task needs.  
Our key insight is to combine complementary views: one summarizes attribute meaning so that descriptive values contribute compact semantic signals, and another captures relational context while assigning greater influence to neighbors that are more informative for the task. 
In \dbname{}, we propose SDEA~\cite{SDEA}, which learns semantically informed representations by combining an attribute view and a relation view.  
The attribute view summarizes values, including long text, into compact vectors using a pretrained language encoder.  
The relation view aggregates signals from related items with selective weighting so that more relevant neighbors have greater influence.  
Although introduced for entity alignment, this design applies more broadly by treating records and linked items as entities and deriving neighborhoods from joins, keys, or time windows.

%% file: sections/5-design-model.tex
\section{Model Layer}
\label{sec:model}
\subsection{Model Storage}
Model Storage in \dbname{} addresses a simple problem: models change while tasks and data evolve, and the system must keep these changes safe, traceable, and efficient. 
It should let the task and data layers read the right model quickly, record updates without heavy cost, and preserve enough context to reproduce any execution. 

Our central insight is to separate fast changing content from stable bases and to let storage be the contract that ties plans, models, and data together.
Small adjustments should be recorded with light metadata, while shared components remain unchanged. Reads should resolve by task intent and schema with clear compatibility checks.
The key idea is a compact, versioned abstraction with four artifact classes: Index, Metadata, Tensor, and Model.
Index supports fast resolution by task and schema. Metadata carries lineage, context, and compatibility so that plans can bind safely. 
Tensor stores intermediate representations that accelerate adaptation. 
Model holds parameters and structure. Updates create new versions by reusing the stable parts and touching only what changes, which keeps refresh and calibration inexpensive and predictable.

\subsection{Construction \& Maintenance}
\subsubsection{Adaptive Model Construction}
Modern data analysis systems serve many datasets, changing tasks, and strict service objectives. 
A model that works well on one dataset often fails to generalize to others, and moving data to external pipelines increases cost and breaks provenance. 
A construction capability inside the data system that adapts models to each workload and data context is therefore necessary. 
The goal of Adaptive Model Construction in \dbname{} is to turn prepared feature views and task specifications into deployable predictors that achieve strong accuracy while respecting limits on time, memory, and latency.

Achieving these aims is challenging even in the structured data setting that dominates relational systems~\cite{RDBMS3,RDBMS4,zeng2024powering}. 
Tables differ in analysis target, missingness, and the feature interactions that drive prediction. 
Traditional models rely on task specific architectures and typically require full retraining for each new table~\cite{xgboost,lightgbm,fttransformer,saint}, which is costly at scale and often brittle under drift. 

A key insight path for \dbname{} is to combine general knowledge with table aware specialization during construction. 
The system should learn a compact descriptor that summarizes table and task signals, compose a small pool of reusable base predictors according to that descriptor, update only the components that matter for the current table, and guide adaptation with an explicit cost model so that quality, latency, and memory remain within budget. 
This approach preserves provenance because construction happens inside the data system, reduces wasted computation by avoiding full retraining, and improves robustness by letting the composition change with the table while the shared knowledge remains reusable.
We implement this view in Adaptive Model Construction and instantiate it with AixelNet~\cite{AixelNet} for structured data prediction task. 
AixelNet learns table level meta representations that relate heterogeneous tables, maintains a compact ensemble of column aware Transformer predictors, uses a lightweight routing network to compose predictors per table, applies sparse updates during pre training to promote specialization and efficiency, and performs short cost aware fine tuning to meet service objectives. 

\subsubsection{Model Drift Detection}
\dbname{} seeks to keep models effective as tasks and data change, while keeping cost low and results traceable.
We detect drift in a task-aware way so that actions are triggered where they matter most.
For example, a forecasting model may keep its overall accuracy, yet errors rise for a single high-impact group;  the system should catch this early and report it clearly.

Our key insight is simple: monitor utility where the task cares about it, not only global similarity.
We track performance on task-relevant slices defined with help from the task plan and feature management (for example, region, product group, or latency tier).
A drift signal is a persistent change on these slices beyond recent variability.
We combine label-based and label-free checks and reuse artifacts already produced by the system.
When labels lag, we watch score and calibration stability on short windows.
When labels arrive, we compute primary metrics per slice (such as AUC or RMSE) and compare them with rolling baselines using lightweight tests.
Input shift summaries from the data layer help explain changes but do not trigger actions on their own.

Each deployed model carries a small monitoring spec that lists slices, metrics, and budgets.
At runtime the system maintains sliding windows, updates statistics in micro-batches, and aggregates them into a single drift score with hysteresis to avoid oscillation.
If the score crosses a threshold, the detector emits an event with the affected slices, metric deltas, and links to small evidence samples stored with the model.
Drift events feed Construction \& Maintenance for follow-up.
Depending on the diagnosis, the system may keep the model as is, refresh calibration, apply a short fine-tuning step, or try a candidate from the model zoo.
A brief summary is also returned to the task layer to inform future planning.
This approach keeps models reliable in dynamic settings without heavy overhead and remains easy to audit and tune.

\subsubsection{Model Update}
Model Update in \dbname{} manages model versions, structures, and parameters so that predictors evolve safely and efficiently as tasks and data change. 
The design is aligned with \dbname{} goals of usability, adaptivity, performance, and extensibility. 
We integrate an internal pipeline versioning module, MLCask~\cite{mlcask}, as a core part of \dbname{}. 
It provides reproducible snapshots, explicit lineage, and artifact reuse across updates, and we extend these ideas from end to end pipelines to fine grained model artifacts and multi tenant workflows.
Every update produces a versioned snapshot with immutable metadata.
Small adjustments such as calibration layers or routing changes are stored as deltas to avoid duplicating large weights. 
Linear updates follow a single lineage and include calibration refresh, short fine tuning, and full retraining.
Nonlinear updates support concurrent contributions from different users or domains.
\dbname{} maintains branches, applies policy driven merges at the smallest differing artifact level, records conflicts and resolutions, and preserves side by side comparability.

Pre trained backbones that serve multiple tasks are updated conservatively.
\dbname{} prefers lightweight adaptations such as adapters, low rank deltas, or routing adjustments for ensemble or MoE style models.
Shared changes are gated by cross task evaluation on a stable panel.
When risk is detected, the system recommends per task adapters instead of global retraining.
To control cost, Model Update reuses selection manifests from the data layer and cached features from the task layer.
This arrangement keeps updates predictable and auditable while supporting rapid refresh for single lineage models and controlled evolution under multi user collaboration.

%% file: sections/6-design-task.tex
\section{Task Layer}
\label{sec:task}

\subsection{Query Parser}
The Query Parser is a robust and task-oriented component of \dbname{}. 
It interprets user intent and converts it into well-typed tasks inside \dbname{} while extracting constraints and quality targets.
The parser balances flexibility for natural language input with predictable structure for downstream optimization and execution.

Therefore, we propose AixelParser, which integrates a Declarative Interface with Task Parsing. 
The declarative interface defines a small, typed schema for objectives, targets, filters, and preferences, and it also accepts natural-language requests. 
Users may provide key–value fields, free-form descriptions, or a mixture of both. 
The interface normalizes values and units, checks domain validity, extracts structured constraints from natural language, and preserves expected outputs while abstracting away execution details. 
The result is a self-contained, verifiable request that is both schema-grounded and language-friendly.

Given a clear declaration, the parser performs intent grounding, constraint extraction.
The parser consolidates the extracted constraints and inferred intent to determine the execution task and the system-internal operation schema.
It selects appropriate operator families, specifies required inputs and outputs, and outlines control-flow requirements. The result is a system-level operation specification that serves as the canonical execution contract for \dbname{} and feeds directly into execution plan construction.

Within the declarative interface of \dbname{}, we design a  NL2SQL interface named GeVeR.
GeVeR maps user intent to relational form at a high level by structuring the request into schema-aware representations, verifying semantic fidelity to the specification, and refining candidates when inconsistencies are detected.
As part of the declarative interface, it enables consistent, schema-grounded interpretation of complex inputs and returns well-formed SQL specifications for downstream execution in \dbname{}.

\subsection{Task Execution}

\subsubsection{Operator Definition}
Operators are the atomic, composable units used by the Task Execution layer to build executable plans.
Each operator encapsulates a well-typed interface, a deterministic function over its inputs, and optimization metadata such as selectivity hints, latency and memory profiles, and cacheability.
Plans are expressed as directed acyclic graphs (DAGs) whose nodes are operators and whose edges carry typed records or tensors. 
This abstraction lets the execution optimizer bind, reorder, parallelize, and tune computations in a principled way.

In \dbname{}, we define three families of operators with consistent contracts.

\begin{itemize}
    \item \textbf{Data Operators}: Interface with the Data Layer. They handle scan, index lookup, filter, project, join, aggregate, sampling, and table slicing, along with other data-related operations. They expose selectivity and cost hints to support pushdown and cost-aware planning.
    \item \textbf{Model Operators}: Interface with the Model Layer. They manage model selection, training, fine-tuning, and monitoring for model drift, as well as other model-centric operations. They report quality and runtime characteristics, supported batch sizes, and version metadata, enabling the optimizer to allocate resources effectively.
    \item \textbf{LLM Operators}: Interface with AI components within the Task Layer to provide LLM-powered assistance. They perform tasks such as task decomposition, plan sketching and structured verification. They provide prompt templates and guardrails, enabling reproducible and budget-aware LLM usage during execution.
\end{itemize}

With these operators, \dbname{} composes reliable plans and enables the Execution Optimizer to bind, reorder, and scale computations to meet latency, quality, and budget targets.

\subsubsection{Execution Plan}
An execution plan is the concrete blueprint that turns a parsed intent into actionable steps in \dbname{}. 
It is central to meeting latency, quality, and budget goals, since plan structure determines data movement, parallelism, and reuse. 
Given a parsed intent, \dbname{} executes in two stages.

\begin{itemize}
    \item \textbf{Operator Binding.} The \dbname{} maps each step of the task specification to concrete operators drawn from the Data, Model, and LLM operator families. Binding selects concrete operators by aligning task semantics and schema bindings with operator capabilities and data locality. Each candidate is scored on expected accuracy and runtime, index availability, batching feasibility, cache reuse opportunities, and historical telemetry, and the best match is bound with its parameters and implementation variant while a safe fallback is retained for runtime adaptation.
    \item \textbf{Plan Synthesis.} Bound operators are composed into a directed acyclic graph with explicit data dependencies. The planner orders and groups operators, inserts pushdown points, selects exchange and materialization boundaries, and annotates the graph with parallelism, caching, and retry policies. The result is an executable plan with hooks for runtime signals to support adaptive re-parameterization.
\end{itemize}

Based on this workflow, we design AixelAsk~\cite{AixelAsk}, an LLM-powered framework for large-table question answering.
The question is first parsed by an LLM to infer intent and produce a task graph, which is then instantiated during Operator Binding with concrete retrieval and reasoning operators.
Retrieval employs row–column ensemble scoring to identify relevant tables to materialize compact evidence subtables, whereas reasoning performs controlled stepwise inference over the assembled evidence.
During Plan Synthesis, the bound operators are assembled into a directed acyclic plan with explicit dependencies, early row and column selection, and parallel execution on independent branches.
The optimizer also combines compatible steps when appropriate to shorten the critical path while preserving answer fidelity. 
This design improves retrieval efficiency and stabilizes reasoning on large tables.

\subsubsection{Optimization}

The execution optimizer is a core component of \dbname{}, responsible for translating decomposed query intents into executable plans while meeting multiple objectives, including latency, answer quality, and monetary cost. 
Because \dbname{} serves multiple tasks, it must synthesize task specific execution plans and apply correspondingly tailored optimization strategies.

Traditionally, optimizers rely on rule based rewrites combined with cost-based search~\cite{seshadri1996cost,freytag1987rule}. 
As the optimization search space grows with heterogeneous operators and multi objective trade offs, learning based approaches have been explored~\cite{yu2022learning-based, zhu2022learned, chen2023loger}. 
Models are trained to learn cost predictors or to directly rank and enumerate plans using supervised learning or reinforcement learning. 
While effective within their training domains, these methods depend heavily on model training and exhibit limited generalization across multiple tasks.

Therefore, we propose AixelOptimizer, a multi task execution optimizer. 
It combines LLM guided semantic priors with a lightweight symbolic layer that validates, tunes, and schedules plans using observed costs and statistics. 
This hybrid optimizer yields task specific plans and balances latency, answer quality, and budget.

In AixelAsk, we design a DAG-optimizer that treats the DAG as the execution blueprint. 
Nodes are scheduled in topological order and independent branches execute in parallel. 
To further improve efficiency, the DAG-optimizer reduces DAG complexity through safe node merging under answer quality constraints. 
It considers two merge patterns, namely compatible siblings that share a common parent and short sequential chains. 
Merges are allowed only when tasks are semantically related and inputs remain within safe limits, ensuring that outputs stay separable and faithful. 
These choices shorten the critical path, reduce redundant calls, and preserve the original reasoning structure.

For LLM tasks that require multiple prompt invocations, we design a Batch Optimizer~\cite{ji2025optimized, cedept-chi-2024}. 
It forms adaptive batches by grouping queries with similar intent and structure, selects shared demonstrations that maximize coverage and alignment, and tunes batch size and template layout to respect context and budget limits. 
It estimates query and demonstration affinity using embeddings with light schema cues, enforces diversity so that demonstrations cover distinct patterns, and removes duplicate prompt segments to reduce overhead. 
It schedules batches to exploit caching and parallel execution while controlling latency, and supports online rebalancing when runtime signals indicate drift or imbalance. 
Integrated into the execution pipeline, the Batch Optimizer lowers per query overhead and API calls, improves throughput, and sustains answer quality across tasks.



%% file: sections/7-conclusion.tex
\section{Conclusion}
Modern analysis needs data and learning to work together under accuracy, latency, and cost limits.
This paper presented \dbname{}, an AI native data analysis system that unifies data, models, and tasks in one execution framework.
The system organizes work across four layers and expresses workflows as operator graphs with clear interfaces and versioned artifacts.
This design keeps plans coherent and results reproducible.
The data layer provides unified indexing, constraint aware discovery, task aligned selection, and feature management for selection, interaction, and embedding.
These functions improve effective data quality and lower cost before heavy computation.
The model layer offers versioned storage, adaptive construction that respects service goals, task aligned drift monitoring, and controlled updates that reuse shared parts.
The task layer compiles intent into operator plans and schedules execution with reuse and caching so that plans meet stated accuracy and latency targets.
Together, this design reduces interaction overhead, supports adaptation to evolving data scenarios, maintains performance within accuracy and cost constraints, and provides strong extensibility.